\documentclass[conference]{IEEEtran}
\usepackage{amsfonts}
\usepackage{amssymb}
\usepackage{latexsym}
\usepackage{cite}
\usepackage[tight,footnotesize]{subfigure}
\usepackage{color}
\usepackage{mathtools}
\usepackage{amsmath}

\usepackage[T1]{fontenc}
\usepackage[]{lmodern}
\usepackage[utf8]{inputenc}
\usepackage[autostyle]{csquotes}
\usepackage{textcomp}

\usepackage[]{siunitx}
\usepackage{textcomp}

\usepackage{todonotes}

\newcommand{\rrn}{r_{r}}
\newcommand{\renz}{r_{\text{enz}}}
\newcommand{\Renz}{R_{\text{enz}}}
\newcommand{\onemicro}{\SI{1}{\micro\meter}}

\newcommand{\ts}{t_s}
\newcommand{\tend}{t_{\text{end}}}
\newcommand{\dist}{\textit{d}}
\newcommand{\hl}{\Lambda_{1/2}}
\newcommand{\hle}{\Lambda^{\SI{1}{\micro\meter}}_{1/2}}
\newcommand{\vlp}{V_{\text{lp}}}
\newcommand{\vtot}{V_{\text{totenz}}}
\newcommand{\erfc}[1]{\text{erfc}\left[ #1 \right]}

\newcommand{\etal}{\textit{et al.}}
\newcommand{\ap}{\textquotesingle \;}

\begin{document}


\title{Effective Enzyme Deployment for Degradation of Interference Molecules in Molecular Communication}

\author{
\IEEEauthorblockN {Yae Jee~Cho, H.~Birkan~Yilmaz, Weisi Guo*, and Chan-Byoung~Chae}\\
       \IEEEauthorblockA{School of Integrated Technology, Yonsei Institute of Convergence Technology, Yonsei University, Korea\\
     *School of Engineering, University of Warwick, UK\\
      Email: \{yjenncho, birkan.yilmaz, cbchae\}@yonsei.ac.kr, *weisi.guo@warwick.ac.uk}}

\maketitle

\begin{abstract}
In molecular communication, the heavy tail nature of molecular signals causes inter-symbol interference (ISI). Because of this, it is difficult to decrease symbol periods and achieve high data rate. As a probable solution for ISI mitigation, enzymes were proposed to be used since they are capable of degrading ISI molecules without deteriorating the molecular communication. While most prior work has assumed an infinite amount of enzymes deployed around the channel, from a resource perspective, it is more efficient to deploy a limited amount of enzymes at particular locations and structures. This paper considers carrying out such deployment at two structures--around the receiver (Rx) and/or the transmitter (Tx) site. For both of the deployment scenarios, channels with different system environment parameters, Tx-to-Rx distance, size of enzyme area, and symbol period, are compared with each other for analyzing an optimized system environment for ISI mitigation when a limited amount of enzymes are available. \\
\end{abstract}


\IEEEpeerreviewmaketitle


\section{Introduction}
The field of nano-devices is gaining more significance, enlarging the influence and potential of nanocommunication in fields such as bioscience, environmental engineering \cite{akyildiz2011nanonetworksAn}. A prominent means for short range communications that has attracted a good deal of research is molecular communication via diffusion (MCvD) \cite{Farsad2016ComSurv, KimNa2013MCvD, nakano2013molecularC, Akka2015SIGREC}.  MCvD uses molecules to convey signals between short-ranged transmitter and receiver nodes, but inter-symbol interference (ISI) increases the error rate. Predicting how the receiver will perceive the ISI-mixed molecular signal is therefore important for analyzing the performance of a MCvD system. Although molecular communication has its' advantages over RF technologies in terms of pathloss in specific environments~\cite{Guo2015MCVSRF}, a crucial drawback is its' signal propagation time and great deal of channel randomness~\cite{Akif2015MCwE, KimNa2014ChaNoise} leading to severe ISI problems. ISI in MCvD is defined as any molecules that survive in the channel beyond the receiver's sampling interval. An obvious way to mitigate ISI is to increase the symbol period~\cite{KimNa2014SymInt}. This, however, will detrimentally decrease the data-rate. A more reasonable mitigation method is to deploy enzymes that can degrade the remaining ISI molecules in the channel.

In \cite{Noel2014MCwE}, Noel \etal{} derived the received number of molecules at a passive and spherical receiver in an enzyme deployed 3-dimensional (3D) channel. Simulations used binary concentration shift keying (BCSK) modulation and improved bit-error-rate (BER) was presented. In \cite{lin2012signal}, Lin \etal{} proposed a quantity-based modulation system where different signal magnitudes correspond to different digital information. Based on the likelihood ratio test and Bayesian criterion, the decision rules at the receiver were obtained, and a decision feedback approach for ISI cancellation was proposed. In \cite{Akif2015MCwE}, Heren \etal{} gave an analytical modelling of an enzymatic 3D MCvD channel with a point transmitter and a spherical absorbing receiver. Enzymes with different degradation rates were considered and their influence was evaluated. Many studies have demonstrated the effectiveness of enzymes on ISI mitigation. They all assume, however, that an infinite amount of enzymes are available within the channel, which is unrealistic. From a resource perspective, like frequency in RF communication, the amount of enzymes is limited and thus should be used effectively. Here, effectively using the enzymes means to have the maximum level of ISI mitigation with a fixed amount of enzymes.

In this paper, we model mainly two different limited enzyme deployment scenarios-- enzymes deployed around the Rx and around the Tx, named in this paper as \enquote{around Rx} and \enquote{around Tx}. The enzymes are assumed to be stationary without degrading themselves in any circumstance. The system is a 3D MCvD channel with a spherical and non-passive transmitter and receiver.The Tx reflects (i.e., molecules that enter the Tx are put back to their previous positions which are outisde the Tx) and the Rx absorbs the molecules (i.e., molecules that enter the Rx are eliminated from the channel after counted). Enzymes are added to the system in an extending, homocentric spherical region around the Rx or Tx. By looking at different channel environments, we investigate the optimal scenario that mitigates ISI.  The channel variation parameters are Tx-to-Rx distance, size of enzyme area, and symbol period.

The remainder of the paper is organized as follows. Section II explains the concept and the mathematical expressions for MCvD channels with enzymes. The section also includes reasons for ISI and the general effect of enzymes on ISI. Section III elaborates on the detailed topology and different scenarios of our limited enzymatic molecular system along with geometric factors to be considered. Section IV discusses the metric used to evaluate ISI and analyzes the results of the simulations for different deployment scenarios. Section V concludes the paper. 

\begin{figure}[t]
\centering{\includegraphics[width=1\columnwidth,keepaspectratio]
{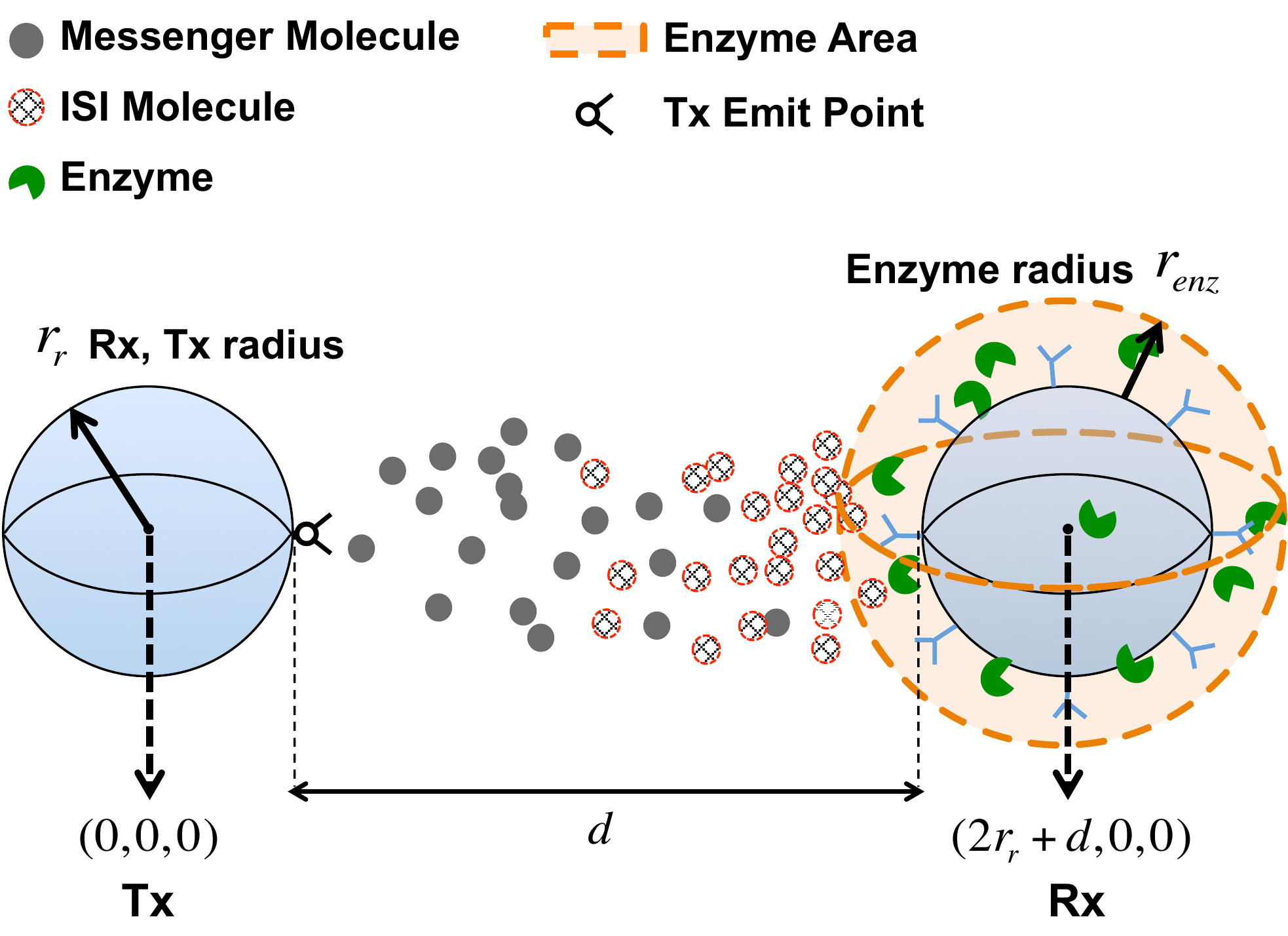}}
\caption{MCvD system model with limited amount of enzymes for \enquote{around Rx} scenario.} \label{fig:MCvDwE_intro}
\end{figure}

\section{System Model}
\subsection{Molecular Communication via Diffusion}
In MCvD, modulation can be done by utilizing different properties: number, type, or time of release of messenger molecules \cite{Farsad2016ComSurv,KimNa2013MCvD}. After modulation scheme is decided, the Tx emits a predetermined number of messenger molecules at every symbol period according to the intended message. To reduce energy consumption, generally zero molecules are emitted for one of the symbols. Once the molecules are released, these act as a signal and propagate through the channel by free diffusion until reaching the receiver as in Fig.~\ref{fig:MCvDwE_intro}. Therefore, the hitting probability of messenger molecules is closely related to the received signal. If nearly all of the messenger molecules arrive at the receiver within their symbol period (i.e., before the next emission), ISI will not occur. Messenger molecules, however, remain and interfere with the following symbols due to the heavy tail nature of the molecules' arrival time histogram.

The ISI problem arises due to the channel randomness that causes a significant amount of molecules to wander about, reaching the receiver after the next emission. Instead of increasing the symbol period with the trade-off of low data-rate, using enzymes to degrade the messenger molecules can be a reasonable solution for ISI mitigation. Degrading late molecules with higher probability than current molecules is a desired situation, which is ensured by the consequences of long-term wandering molecules in the enzyme region. The trade-off of using enzymes is the loss of signal power, which can be compensated by lowering the threshold for decoding. 

A molecule's hitting rate to the receiver for time $t$ in a non-enzymatic channel with an absorbing 3D receiver and a point Tx is formulated as follows:
\begin{align}
h(t) = \frac{\rrn}{\dist+\rrn}\frac{\dist}{\sqrt{4\pi Dt^3}}e^{-\frac{\dist^2}{4Dt}}  
\label{hitp}
\end{align}
where $\rrn$, $\dist$, and $D$ is the receiver radius, the Tx-to-Rx distance, and the diffusion coefficient, respectively~\cite{Yilmaz20143Drx}. Then the fraction of molecules received until time $t$ will be the integral from $0$ to $t$ as
\begin{align}
F(t) = \int\limits_{0}^{t}h(t^{\prime})dt^{\prime}  = \frac{\rrn}{\dist+\rrn}\erfc{\frac{\dist}{\sqrt{4Dt}}}. 
\label{hitn}
\end{align}
Channel dynamics of a 3D channel with a point Tx and an absorbing spherical Rx are determined by \eqref{hitp} and \eqref{hitn}. They give a general understanding of the system but do not perfectly match the system researched in this paper since our system has a reflecting spherical Tx and also enzyme effect is considered. The spherical factor of the Tx and enzyme effect are applied by adding the corresponding features to the simulation accordingly.

\vspace{2mm}
\subsection{Enzyme Dynamics}
In terms of chemical reactions, enzymes are critical. They speed up reactions to the level where mechanisms can function properly. The main function of enzymes is to decompose certain substrates with specificity, such as acting on particular types of molecules or chemical bonds. 

The enzyme chemical reaction is defined as:
\begin{align}
	E+S  \xrightleftharpoons[k_{-\!1}]{\,k_1\,} ES \xrightarrow{k_p} E+P
\label{enzm_ch}
\end{align}
where $E, \, S,\, ES, \,P$, and $k_{-\!1}$ and $k_1$ is the enzyme, substrate, enzyme-substrate compound, product, and rate of reactions respectively.  
Under the assumption of a fast enzymatic reaction \cite{Akif2015MCwE}, that is, an exponential decaying degradation, the concentration of the messenger molecules (substrate) at time $t$ and the initial substrate concentration $C_0$, will be an exponential decay function as
\begin{align}
	C(t)        &= C_0e^{-\lambda t} 		\label{prob_decay} \\
	\lambda &= [S][E] = \frac{\ln(2)}{\hl}            \label{deg}
\end{align}
where $\lambda,\,\hl$ corresponds to the degradation rate and half-life of the enzyme respectively. For $C_0 = 1$, \eqref{prob_decay} becomes the probability of degradation for each messenger molecule at each step of its\ap movement. Now we can apply enzymes to the MCvD mathematical analysis \eqref{hitp} and \eqref{hitn}.

Assume that the arrival of messenger molecules at time $t$ to the receiver is $f_A(t)$, and the probability for degradation time $T$ being greater than arrival time $t$ is $P_B (T>t).$ Then the probability of messenger molecules hitting the Rx before degradation is
\vspace{2mm}
\begin{align}
h(t | \lambda) &= f_A(t)\cdot P_B(T>t) \\
&= \frac{\rrn}{\dist+\rrn}\frac{\dist}{\sqrt{4\pi Dt^3}}e^{-\frac{\dist^2}{4Dt}-\lambda t}.
\end{align}
The total number of received molecules until time $t$ becomes
\begin{align}
\label{hiten}
\begin{split}
 F(t|\lambda)  &=\int\limits_{0}^{t}h(t^{\prime} | \lambda)dt^{\prime} \\
  		     &= \frac{1}{2}\frac{\rrn}{d\!+\!\rrn} \left\{e^{-\dist \sqrt{\frac{\lambda}{D}}} \,\erfc{\frac{\dist}{\sqrt{4Dt}} -\sqrt{\lambda t}} \right. \\
  	             &+ \left. e^{\dist\sqrt{\frac{\lambda}{D}}} \,\erfc{\frac{\dist}{\sqrt{4Dt}} +\sqrt{\lambda t}}\right\} \,
 \end{split}
 \end{align}
for the point source scenario where enzymes are deployed everywhere. 

\begin{figure}[t]
\centering{\includegraphics[width=1\columnwidth,keepaspectratio]
{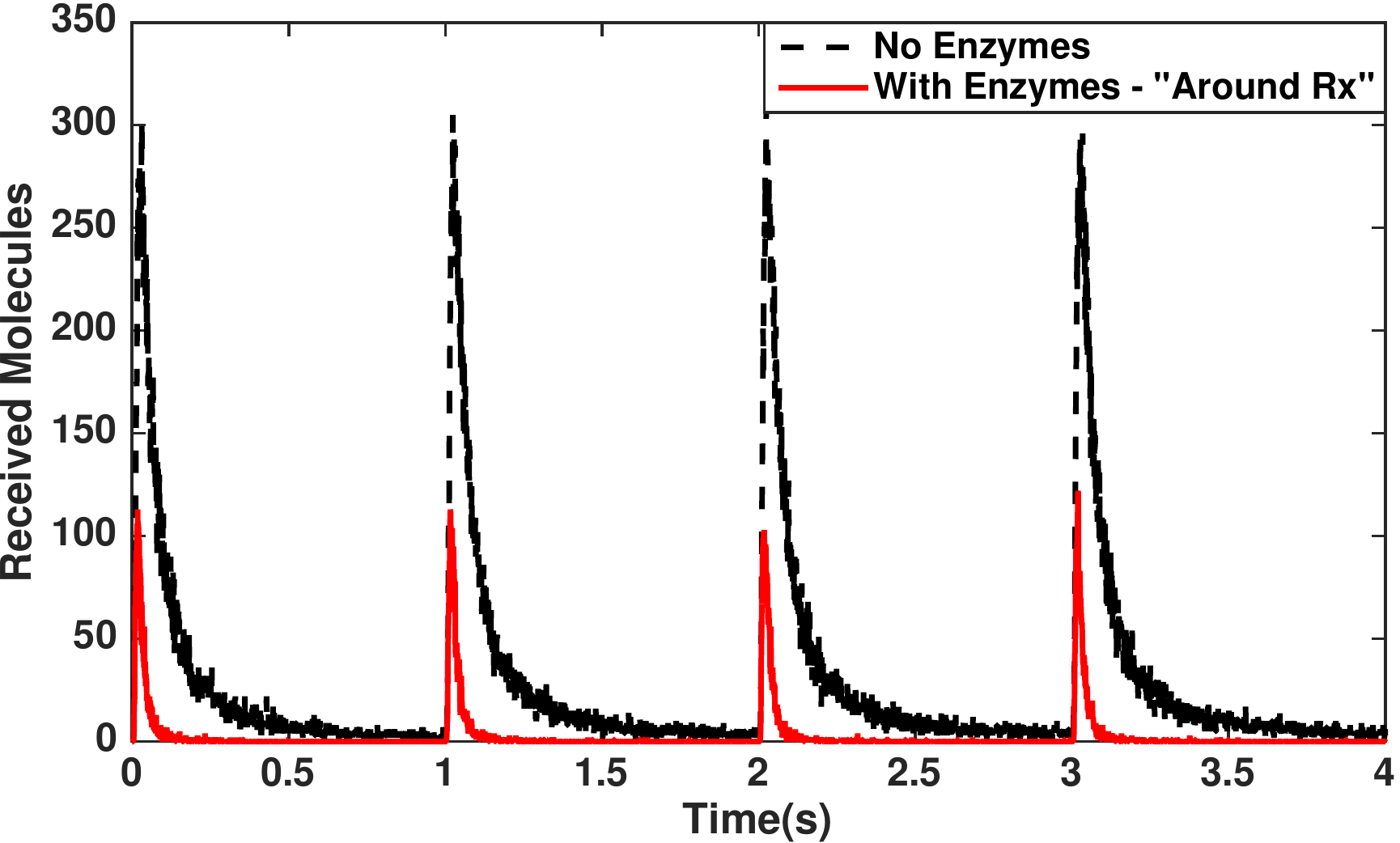}}
\caption{Received signal for the \enquote{around Rx} system with four symbol periods. \text{(${\dist=\SI{4}{\micro\meter}}$, ${\rrn=\SI{5}{\micro\meter}}$}, \text{${\renz=\SI{4}{\micro\meter}}$, ${\ts=\SI{1}{\second}}$, ${\tend=\SI{4}{\second}}$, ${\hl=\SI{0.002}{\second}}$).}}  \label{fig:No_With_Enz_4ts}
\end{figure}

To first see the enzymes' effect on the channel in general without considering a specific type of deployment, Fig.~\ref{fig:No_With_Enz_4ts} shows the received signals for systems where enzymes are deployed (\enquote{around Rx}) and not deployed. Other system parameters besides the presence of enzymes were set as identical. Very clearly the tail part (i.e., ISI) of the \enquote{With Enzymes} signal decreased significantly, looking almost like an impulse function, compared to that of \enquote{No Enzymes}. Hence, while it is verified that enzymes do reduce the ISI, how to use the enzymes in a resource-efficient manner is yet unclear. 


\section{Enzyme Deployment Scenarios}
\subsection{Topology}
The most probable enzyme deployment locations that are easy to implement and may effectively mitigate ISI is around Rx and around Tx. The difference between the two is \textit{when} degradation is being done: when molecules are emitted or close to the receiver. The two scenarios are compared in ISI perspective. They may be extended to other structures that are easy to implement such as a random deployment strategy~\cite{Cho2016ENZMIT}. 

Depicted in Fig.~\ref{fig:MCvDwE_intro}, our system's Rx and Tx are spherical with same $\rrn$. The Rx absorbs the hitting molecules by eliminating them from the channel after counting while the Tx reflects them by placing them back to their original positions. Such concept of reflectivity in MCvD is also applied to other studies such as \cite{Felice2015RFLTX}. At the start of each symbol period, the Tx either does not emit or emit messenger molecules towards the receiver at the \textquotedblleft Tx Emit Point\textquotedblright\, in Fig.~\ref{fig:MCvDwE_intro}. The Rx counts the number of molecules that are absorbed constituting the received signal for demodulation. This paper does not consider in specifics regarding on which type of demodulation will be used, since the focus is rather on how the differently deployed enzymes will affect the shape of the received signals.

\vspace{2mm}
\subsection{Enzyme Channel Structure}

\begin{figure}[t]
\centering{\includegraphics[width=0.95\columnwidth,keepaspectratio]{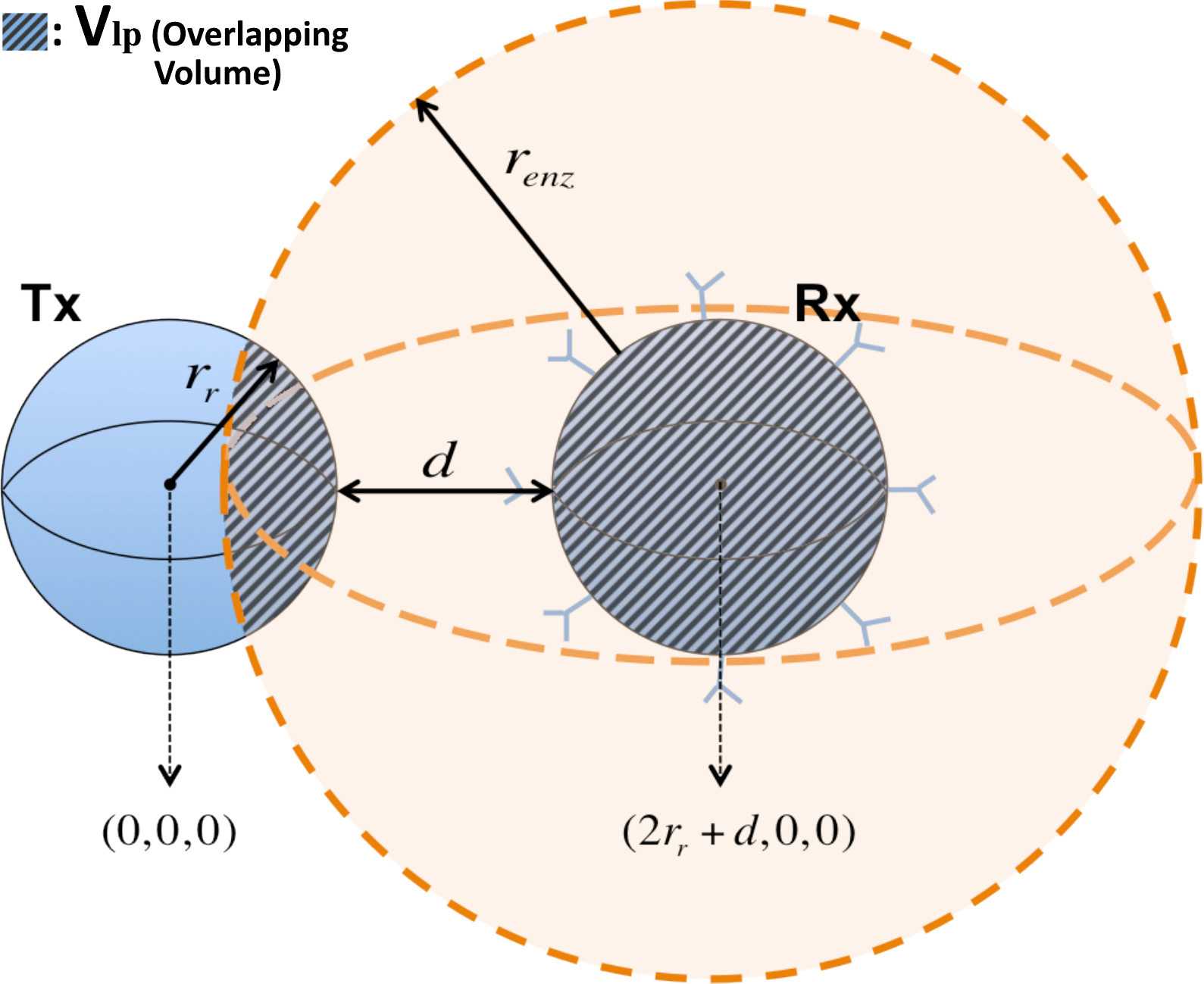}}
\caption{Geometry of overlapping volume for \text{$\dist \!<\! \renz \!<\! \dist + 2\rrn$.}} \label{fig:overlap_geo}
\end{figure}

The enzymes are deployed to only the enzyme area with an extension of $\renz$ as shown in Fig.~\ref{fig:MCvDwE_intro}. Since the degradation probability applies to only the molecules inside the enzyme area, a virtual fixed spherical enzyme region is maintained. Moreover, a fixed unit amount of enzymes are identically deployed for every scenario by the method mentioned below. 

From \eqref{deg} the enzyme concentration is defined as, $[E]=\ln2/ (\hl[S])$. Hence, with $[E]\simeq1/\vtot$ maintained for each of the different scenarios, where $\vtot$ is the total volume of the enzyme region, an equal unit enzyme amount of 1 for every different enzyme deployment scenarios is achevied. Note that the number of enzymes are kept constant, not the concentration. This method is applied to the half-life by setting an effective half-life explained in the subsequent part of this paper.

The enzyme area is an extending spherical area to the Rx or Tx with volume $\dfrac{4}{3}\pi(\rrn+\renz)^3-\vlp$. $\vlp$ is the volume of Rx and Tx that overlaps with the enzyme area, which needs to be excluded from the total enzyme area volume. This is because the enzymes are not employed within the Rx or Tx which makes the overlapping area a non-enzyme area. $\vlp$ changes according to the $\renz$ value. When $\renz \geq \dist+ 2\rrn, \, \vlp$ simply becomes the volume of both Tx and Rx. When $\dist+2\rrn > \renz > \dist, \, \vlp$ becomes the volume of either Tx or Rx plus the lens-similar shape \cite{Kern1948sph2sph}, which is the intersection of two spheres ($\vlp$ in Fig.~\ref{fig:overlap_geo}). Lastly when $\renz \leq \dist, \, \vlp$ will just be the volume of a single Tx or Rx. Hence we can consider three cases for $\vlp$ as follows: 
\vspace{2mm}
\begin{align}
 \vlp &= \begin{cases}
 			\dfrac{8}{3}\pi\rrn^3\,, \qquad \qquad \qquad \qquad \; \; \; \; \text{if } \renz \geq \dist + 2\rrn \\[9pt]
 			\dfrac{4}{3}\pi\rrn^3\,, \qquad \qquad \qquad \qquad \; \; \; \; \text{if } \renz  \leq \dist \\[9pt]
 			A + \dfrac{\rrn^2 - \renz^2}{4\dist_c}+ \dfrac{4}{3}\pi\rrn^3\,, \qquad   \text{otherwise}
 		\end{cases} \\
 	A &= \pi(\renz-\dist)^2\,\dfrac{\dist_{c}^2 + 2\dist_c\rrn -3\rrn^2 + 2\dist_c\Renz }{12\dist_{c}}\,, \nonumber \\
 	\dist_c & = 2\rrn + \dist \,, \quad \quad \Renz   = \rrn + \renz \,. \nonumber
\end{align}
Then the total enzyme volume $\vtot$ becomes 
\begin{align}
\vtot = \dfrac{4}{3}\pi \Renz^3 - \vlp \,.
\end{align}

\vspace{2mm}
\subsection{Effective Half-life}
For different $\renz$, the total enzyme volume changes and $\hl$ has to be changed for maintaining a constant number of enzymes. The concentration of enzyme, [E], is inversely proportional to $\vtot$ which is proportional to $\renz$. Hence, from \eqref{deg}, $\hl$ is also proportional to $\renz$. 

Assume a system where $\renz=r_i$, half-life is $\hl^{r_i}$ with total enzyme area $V_{\text{totenz}, r_i}$. Then with two different $\renz$, $\hl^{r_i}$ can be evaluated as
\begin{align}
\hl^{r_2} = \hl^{r_1} \frac{V_{\text{totenz}, r_2}}{V_{\text{totenz}, r_1}}. \label{hl_eq}
\end{align}
Using \eqref{hl_eq}, we calculate the half-life when $\renz =\SI{1}{\micro\meter}$ for different systems and calculate any $\hl^{\renz}$ by multiplying ${V_2}/{V_1}\;$to it, as in \eqref{effective_hl}. This way the effective half-life is changed every time $\renz$ is changed so that the amount of enzymes is kept constant for different scenarios.
\vspace{3mm}
\begin{align}
\hl^{\renz} = \hle \frac{V_{\text{totenz}, \renz}}{V_{\text{totenz}, \onemicro}}. \label{effective_hl}
\end{align}

\vspace{4mm}
\subsection{Simulation System}
For each time frame $\Delta t$,  every molecule moves by diffusion dynamics governed by Gaussian distribution at each dimension \cite{Birk2014SimStudy}, as follows
\vspace{1mm}
\begin{align}
\begin{split}
\Delta\vec{r} &= (\Delta x, \,\Delta y, \,\Delta z)  \\
\Delta x &\sim \mathcal{N}(0, \, 2D\Delta t) \\
\Delta y &\sim \mathcal{N}(0, \, 2D\Delta t) \\
\Delta z &\sim \mathcal{N}(0, \, 2D\Delta t)
\end{split}
\end{align}
where $\Delta\vec{r}$, $\Delta x$, $\Delta y$, and $\Delta z$ correspond to the displacement vector, and the displacements at $x$, $y$, and $z$ dimensions at a time frame of $\Delta t$ and $\mathcal{N}(\mu, \sigma^2)$ corresponds to the Gaussian distribution with mean $\mu$ and variance $\sigma^2$.

Substituting the effective half-life \eqref{effective_hl}, into \eqref{prob_decay}, we get the final probability of not decaying for each $\Delta t$ step for one messenger molecule inside the specified enzyme region as \eqref{prob_decay_renz}.  Now we formulate a degrading function for limited enzymes in a specified enzyme area. 
\begin{align}
\mathbf{P} (\mbox{no degradation}\,|\, \hl^{\renz}) = e^{-\frac{\ln(2)}{\hl^{\renz}}\Delta t} = \frac{1}{2^{\Delta t / \hl^{\renz}}}. \label{prob_decay_renz}
\end{align}

At each $\Delta t$, the molecules that try to enter the Tx are put back to their previous locations. Then, the molecules in the Rx are counted as received, and then removed from the channel. The next step is checking for degradation. For each molecule, \eqref{prob_decay_renz}, the probability for not degrading, is compared to a uniformly distributed random number, and degradation is done accordingly. The remaining molecules continue propagation. This process is repeated until $\tend$, the simulation end time. 


\section{Results and Discussion}
\subsection{Parameters and Performance Metrics}
For each simulation type, 50 replications were done. In our simulations 10 different $\renz$ values were considered from $2$ to $\SI{20}{\micro\meter}$. For every different $\renz$, the effective half-life was calculated and different $\mathbf{P} (\mbox{no degradation}\,|\, \hl^{\renz})$ was applied to the system. Note that when we increase $\renz$ the probability of entering to the enzyme region increases while the enzyme effectiveness decreases.

For the evaluation of ISI, this paper uses the interference-to-total-received molecules (ITR) metric. For a certain symbol period $\ts$, and simulation end time $\tend$, the ITR is defined as:
\begin{align}
\text{ITR}(\ts, \tend) = \frac{F(\tend) - F(\ts)}{F(\tend)} 
\end{align}
where $F(\cdot)$ indicates the total number of molecules received until time $t$. The parameter indicates the portion of ISI molecules to the total messenger molecules received. Note that the ISI molecules are separated from the desired molecules by setting a reasonable value of $\ts$. The molecules received before and after the $\ts$ is each the desired and intergerence signal. In our case, the smaller the ITR, the better the ISI mitigation. In Table~\ref{tbl_system_parameters}, the system parameters and their values or ranges used for the simulations and performance analysis are presented. The parameters used are feasible for actual biological systems with horomones which have a diffusion coefficient around $\SI{100}{\square\micro\meter\per\second}$~\cite{Farsad2016ComSurv}. Moreover, the specified parameters regard a micro-scale cell-to-cell communicatio like the communication between a neuron and a muscle cell.

\begin{figure}[t]
\centering{\includegraphics[width=0.95\columnwidth,keepaspectratio]
{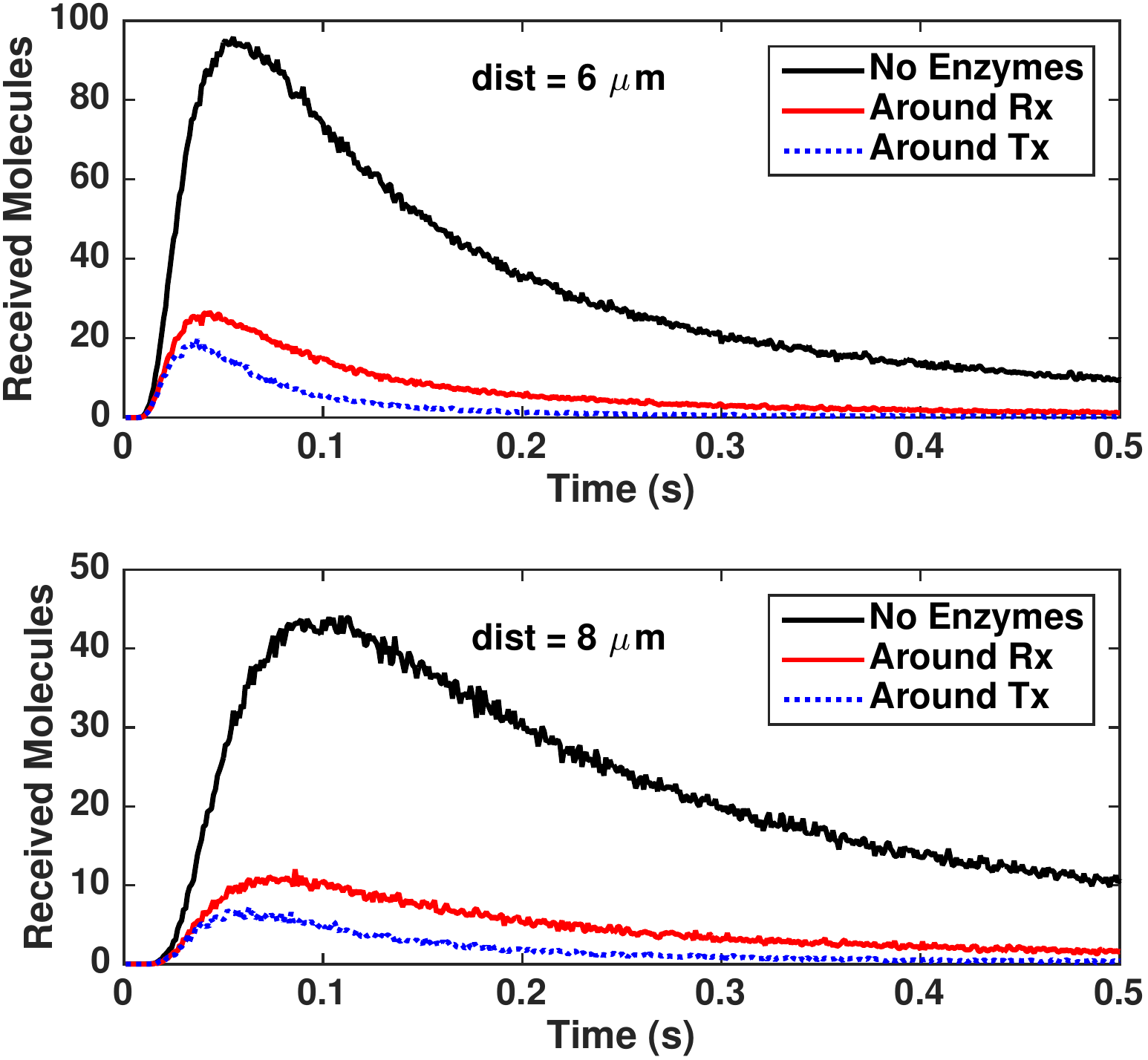}}
\caption{Received signals for \enquote{no Enzyme/around Rx/around Tx} scenarios for distance~\text{$\SI{6}{\micro\meter}\,$(Top) and $\,\SI{8}{\micro\meter}$}(Bottom) with time resolution of \text{$\SI{1}{\milli\second}$ ($\renz = \SI{2}{\micro\meter}$, $\hle = \SI{0.002}{\second}$).}} \label{fig:RxVSTx_Sig}
\end{figure}

\begin{table}[t]
\begin{center}
\caption{Range of parameters used in simulations}
\renewcommand{\arraystretch}{1.1}
\label{tbl_system_parameters}
\begin{tabular}{p{4cm} l}
\hline
\bfseries{Parameter} 							& \bfseries{Value} \\ 
\hline 
 Diffusion Coefficient ($D$) 		& $\SI{100}{\square\micro\meter\per\second} $ \\ 
 Radius of the Rx/Tx ($\rrn$)   		& $\SI{5}{\micro\meter}$         			        \\
 Enzyme Radius ($\renz$)   		& $2 \sim \SI{20}{\micro\meter}$    \\ 
 Distance ($\dist$)				& $4, 6, \SI{8}{\micro\meter}$\\
 Molecules Emitted for one $\ts$	& $5 \times 10^4$molecules\\ 
 Symbol Period ($\ts$)			& $0.5, 0.75, \SI{1.0}{\second}$\\
 Simulation End Time ($\tend$)		& $\SI{2}{\second}$\\
 Unit Half Life ($\hle$)			& $0.002 \sim \SI{0.008}{\second}$\\
 Simulation Step ($\Delta t$)	 	& $\num[retain-unity-mantissa = false]{1e-5} \, \si{\second} $\\
 Replications for Simulation		& 50\\
 \hline
\end{tabular} 
\end{center}
\end{table}

\subsection{Received Molecular Signal}

To distinguish the differences between the enzyme deployment scenarios, the received molecular signals for each scenarios are compared in Figure~\ref{fig:RxVSTx_Sig}. The three signals for \enquote{no enzymes}, \enquote{around Rx} and \enquote{around Tx} scenarios are presented in the figure. Although it is not yet clear how the ITR may appear, the differences in the tail of Tx and Rx signals are apparent for both distances $\SI{6}{\micro\meter}$ and $\SI{8}{\micro\meter}$ in Fig.~\ref{fig:RxVSTx_Sig}. \enquote{around Tx} signals' tail and peak are lower than those of \enquote{around Rx}. The consistent discrepancy in both distances between the scenarios indicates that there is a noticeable difference. Moreover the higher peak and received number of molecules for shorter distance $\SI{6}{\micro\meter}$ indicates that more molecules arrive for shorter range communications. Thus the threshold must be adjusted accordingly when decoding for different distances.

\subsection{ITR Analysis}

\begin{figure}[t]
\centering{\includegraphics[width=0.95\columnwidth,keepaspectratio]
{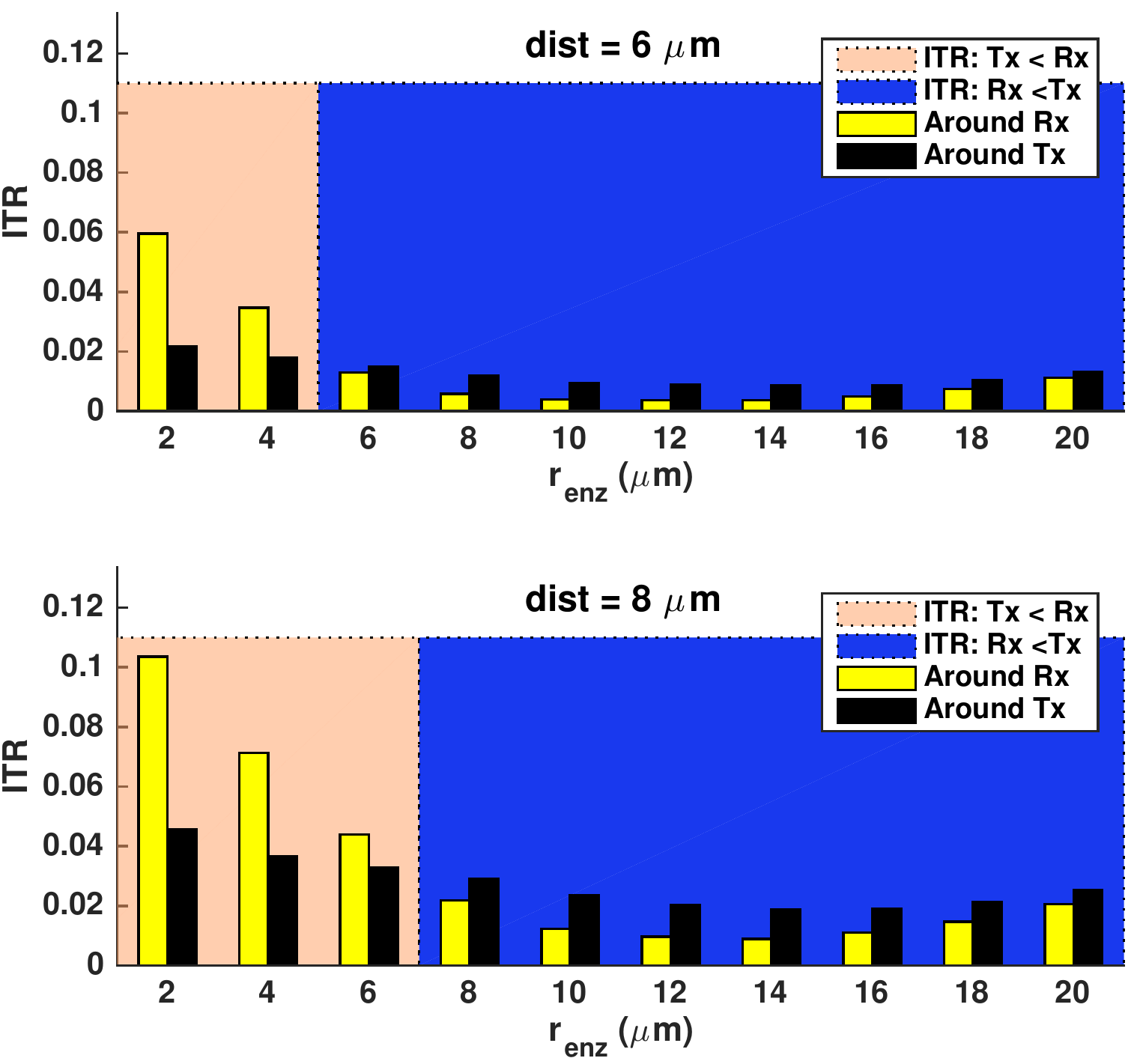}}
\caption{ITR comparison for different \text{$\dist$} and \text{$\renz$ ($\ts = \SI{1.0}{\second}, \, \tend = \SI{2.0}{\second},  \, \hle = \SI{0.002}{\second} $).}} \label{fig:RxVSTx_dist_ITR}
\end{figure}

\begin{figure}[t]
\centering{\includegraphics[width=0.95\columnwidth,keepaspectratio]
{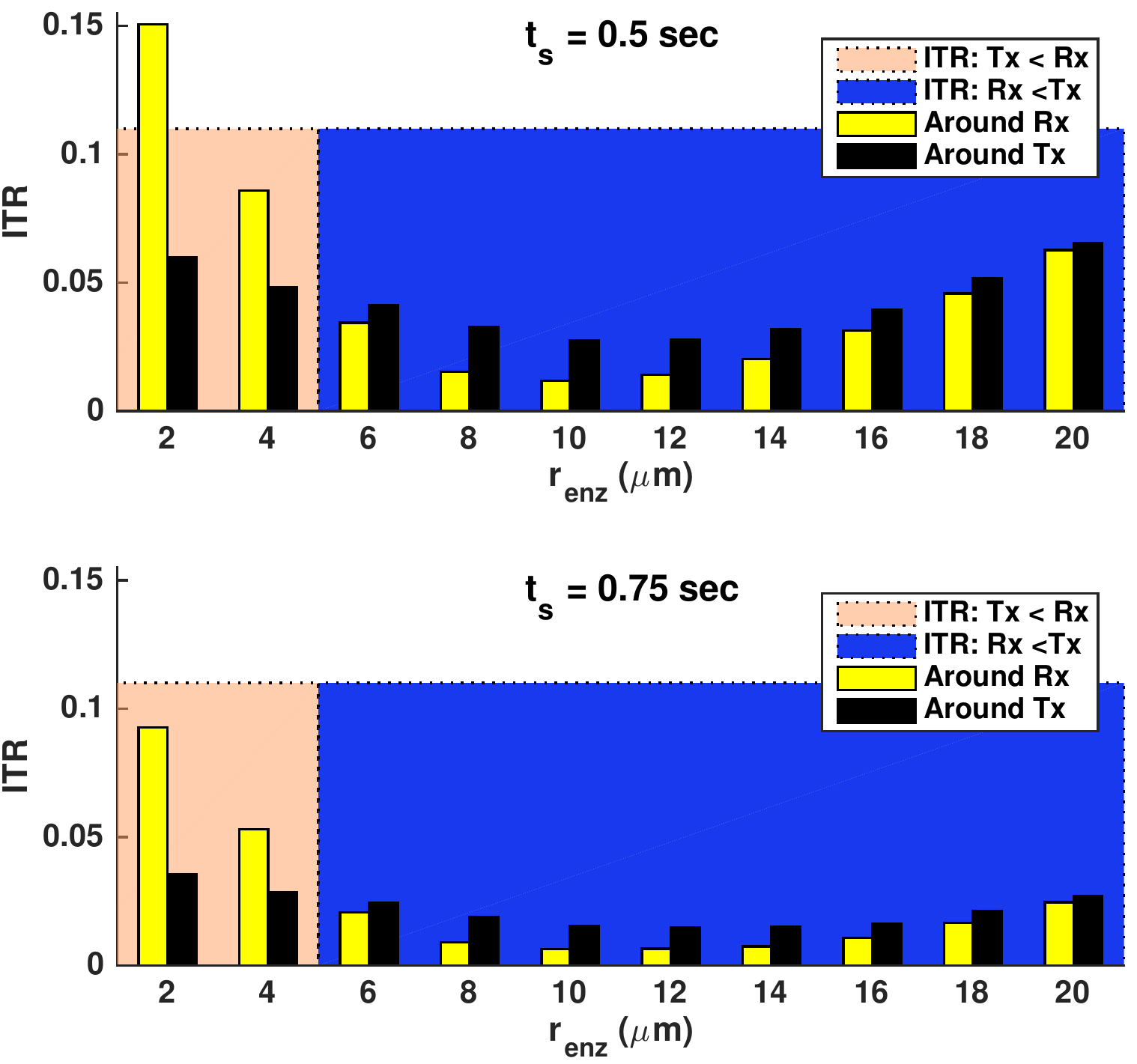}}
\caption{ITR comparison for different \text{$\ts$} and \text{$\renz$ ($\dist = \SI{6}{\micro\meter}, \, \tend =\SI{2.0}{\second}, \, \hle = \SI{0.002}{\second} $).} }\label{fig:RxVSTx_ts_ITR}
\end{figure}

For an in-depth understanding on the difference between the scenarios, ITR analysis is done. Since results from several different channel environments for higher validity is prefered, ITR for fixed $\ts$ and varying $\renz$ is shown in Fig.~\ref{fig:RxVSTx_dist_ITR} for different $\dist$s. The figure shows that there is a limit until when \enquote{around Tx} has a lower ITR than \enquote{around Rx} (shaded lightly). However once that value is exceeded, \enquote{around Rx} has a lower ITR than \enquote{around Tx} (shaded dark). This indicates that when we wish to use a tight enzyme area with a small $\renz$ value, deploying \enquote{around Tx} is better to a certain extent for ISI mitigation. Once $\renz$ exceeds the certain value and the enzyme area gets larger, \enquote{around Rx} deployment mitigates the ISI better. Moreover, notice that the lowest ITR occurs when the enzymes are deployed \enquote{around Rx}, meaning that optimal ISI mitigation occurs for deployment \enquote{around Rx}. The ITR grows as $\dist$ increases, since ISI for long range communication is severe.

Similarly, Fig.~\ref{fig:RxVSTx_ts_ITR} shows the ITR for fixed $\dist$ and varying $\renz$ for different $\ts$s. Figure \ref{fig:RxVSTx_ts_ITR} illustrates the same results with Fig.~\ref{fig:RxVSTx_dist_ITR}. For both Fig.~\ref{fig:RxVSTx_dist_ITR} and Fig.~\ref{fig:RxVSTx_ts_ITR}, note that the ITR does not strictly decrease as the $\renz$ increases. This is because as the enzyme area gets considerably larger, the concentration of the enzymes gets infinitesimal since we deployed a constant amount of enzymes. Hence, the effect of the enzymes on molecule degradation becomes almost negligible, explaining the eventual increase of ITR.


\section{Conclusion}
This paper analyzed different scenarios in which a limited amount of enzymes can be effectively deployed for ISI mitigation. While most prior works have assumed an infinite amount of enzymes, our work took a practical perspective and limited the amount of enzymes into specific enzyme areas. Two cases were considered: limited enzymes deployed \enquote{around Rx} and \enquote{around Tx}. Data showed, for a tight enzyme area, \enquote{around Tx} had a lower ITR than \enquote{around Rx}. However, when the enzyme area exceeded a certain size, \enquote{around Rx} had lower ITR. The lowest ITR occured for \enquote{around Rx}, meaning that maximum ISI mitigation can be done when enzymes are deployed around Rx. Hence, when the $\renz$ size need not to be limited to any value, using \enquote{around Rx} scenario is desirable. These results were valid for all of different $\dist$ and $\ts$. Shorter $\dist$ and longer $\ts$ had a smaller ITR for all the scenarios, meaning the short-distance and long symbol period system results in less ISI. This, however, have the trade-off of lowering data-rate and limiting applicable systems. Further research may be done on analyzing the mathematical and detailed reasonings for the paper's results and also considering different system parameters such as a point Tx or non-equal $\rrn, \; \renz$ situations. Another possible research field would be the enhancements of molecular MIMO system \cite{Bon2016MIMO} with introducing the enzymes \enquote{around Rx} receive antennas. 

\section*{Acknowledgment}
This research has been supported by the MSIP (Ministry of Science, ICT and Future Planning), Korea, under the ``ICT Consilience Creative Program" (IITP-R0346-16-1008) supervised by the IITP (Institute for Information \& Communications Technology Promotion) and by the Basic Science Research Program (2014R1A1A1002186) funded by the MSIP, Korea, through the National Research Foundation of Korea. 

\vspace{0.5cm}

\bibliographystyle{IEEEtran} 

\bibliography{nanocomRefs}

\end{document}